\documentclass[reprint, showpacs,preprintnumbers,amsmath,amssymb,superscriptaddress]{revtex4-1}
\usepackage[utf8]{inputenc}
\usepackage{epsfig}
\usepackage{epstopdf}
\usepackage{amsmath,amssymb,amsthm}
\usepackage{dcolumn}
\usepackage{bm}
\usepackage{float}
\usepackage{graphicx}
\usepackage{subfig}
\usepackage[justification=RaggedRight]{caption}
\usepackage[]{natbib}
\usepackage{hyperref}

\begin{document}

\title{Multiple surface plasmons on an unbounded quantum plasma half-space}
	\author{D. I. Palade}\email{dragos.i.palade@gmail.com}
	\affiliation{National Institute of Laser, Plasma and Radiation Physics,
		PO Box MG 36, RO-077125 M\u{a}gurele, Bucharest, Romania}
	\affiliation{Faculty of Physics, University of Bucharest, Romania}


	\begin{abstract}
	The propagation of surface plasmons on a quantum plasma half-space in the absence of any external confinement is investigated. By means of Quantum Hydrodynamic Model in the electrostatic limit it is found that the equilibrium density profile is a smooth continuous function which, in the linear regime, supports multiple non-normal surface modes. Defining a spectrum function and using a cutting condition, the dispersion relations of these modes and their relevance for realistic dynamics are computed. It is found that the multiple surface plasmons present a significant red-shift with respect to the case of fully bounded quantum plasmas.
	\end{abstract}

	\maketitle

	In the recent years, surface phenomena on a quantum plasma half-space (QPHS) as surface plasmon (SP) or surface plasmon-polariton (SPP) waves have been intensively studied. The interest is driven by their relevance for the next generation of high frequency quantum electronic devices \cite{maier} or other quantum plasma systems as dense astrophysical objects \cite{0953-8984-14-40-307,0034-4885-69-9-R03}, laser-solid based plasma \cite{PhysRevE.75.026404}, metallic nano-structures \cite{scholl2012quantum}, etc. From a theoretical point of view, the SP and SPP on a QPHS are usually investigated within the Quantum Hydrodynamic Model \cite{PhysRevB.78.155412} (QHM). Lately, a lot of progress has been done in this field \cite{RITCHIE1966234,LAzar,Liet,BC,AbdelAziz2012169,Zhu20131736,Moradi} (and the reference therein). 
	
	Despite the impressive amount of work, almost all the existing studies have used a fundamental assumption: the QPHS was considered entirely bounded by a plane resulting in a step-profile of the density. Although not usually mentioned, this choice (found in literature as hard-wall-boundary-conditions) is mainly motivated by the analytical simplifications induced in the linearized QHM. Physically, such a constrain leads to an infinite work function for the electrons in the system which is unrealistic, unless, there is an infinite external confining potential with step profile at the surface of the QPHS. Otherwise, the quantum nature of the electrons should exhibit tunnelling effects, being allowed to have a smooth density profile and consequently a decaying spill-out beyond the ionic volume.
	
	In the present work it is investigated the SP phenomenon in an \emph{unbounded} QPHS taking into account the continuous spatial profile of the density in the absence of any external potential. Thus, the existing approaches are not labelled as unrealistic, but rather it is pointed out that they treat the case of a \emph{bounded} (fully confined) QPHS, while this paper is concerned with the other extrema of no external confinement.
	
	Previous attempts have been made towards the study of electron charge distribution effects on the dispersion relation of a surface plasmon on metal surfaces. We mention the early work of \citet{Bennet} which has considered a simple hydrodynamic model with the assumption of linear surface profile to prove the existence of multiple surface modes. Also, \citet{PhysRevB.3.220} shown that in the frame of Hartree-Fock theory, the frequency of the SP is insensitive to the electron charge profile in the long-wavelength limit. More recently, it was proven the ability of the full QHM to describe realistic quantum plasma profiles and dynamic phenomena in nano-particles \cite{PaladeC60,toscano2015resonance} and metal-surfaces \cite{davidsp}. \citet{PhysRevB.91.115416} presents a comprehensive discussion on the linearized QHM for inhomogeneous electron gas.
	
	The QPHS is defined in an $Oxyz$ Cartesian coordinate system, considering the immobile ions (due to their large inertia) to form a homogeneous background of positive charge with density $n_i$ in the $z>0$ half-space. The system is completed by the electron fluid which is free to occupy the entire space with whatever density is fit to reproduce the ground-state equilibrium. The surface of the quantum plasma is defined as the $z=0$ plane in agreement with equilibrium solutions presented bellow. The dynamics obeys by the QHM equations \cite{PhysRevB.78.155412}:
	\begin{eqnarray}\label{eq_1}
		&\partial_t n+\nabla(n\mathbf{u})  =  0\\
		&(\partial_t+\mathbf{u}\nabla)\mathbf{u} = -\nabla(\varphi+g[n]-\frac{H^2}{2}\tfrac{\nabla^2n^{1/2}}{n^{1/2}})\label{eq_2}
	\end{eqnarray}	
	For legibility, the $(\mathbf{r},t)$ arguments have been dropped for all quantities being automatically implied. We identify the density of electrons $n$, the associated velocity field $\mathbf{u}$, the electrostatic potential $\varphi$ subject to Poisson equation $\nabla^2\varphi=1-n$ and $g[n]=\tau[n]+v_{xc}[n]$ a local functional of density that reproduces the density of kinetic energy together with the exchange-correlation potential. Also, the following scaling has been used: $t\to t\omega_p^{-1}$,  $\mathbf{r}\to\mathbf{r}v_F/\omega_p$, $n\to n n_i$ and $\mathbf{u}\to\mathbf{u}v_F$ where $n_i$ is the ionic density, $\omega_p=e^2n_i/m\varepsilon_0$ is the plasma frequency and $v_F=\hbar/m(3\pi^2n_i)^{1/3}$ defines the Fermi velocity. With these, the scaled Planck constant $H=\hbar\omega_p/mv_F^2$ is obtained.
	
	The ground-state of the quantum plasma is the equilibrium solution ($\partial_t\equiv 0$)  of the QHM which minimizes the total energy. The latter condition imposes $\mathbf{u}=0$, therefore, the momentum eq. \ref{eq_2} becomes a partial differential equation for density. Defining by a Madelung transform the pseudo-wave function $\psi=\sqrt{n}$, one can rewrite\cite{PhysRevB.78.155412,PaladeC60} the stationarity condition as a self-consistent non-linear Schrodinger equation:
	\begin{equation}\label{eq_3}
		(-\frac{H^2}{2}\nabla^2+\varphi+g[n])\psi=\mu\psi
	\end{equation}
	where $\mu$ can be interpreted as Fermi level. This reformulation in an eigenvalue problem has some numerical advantages. Most important, the Bohm potential $n^{-1/2}\nabla^2n^{1/2}$, which is singular at $n=0$, has been transformed in a simple diffusion term $\nabla^2\psi$. Although the resulting operator $\hat{h}=-H^2/2\nabla^2+\varphi+g[n]$ is self-consistent with the solution, it acts linearly on $\psi$ and, consequently, it can be diagonalized with standard techniques. Moreover, the spill-out of density beyond $z=0$ is obtained naturally from the smoothness of the eigenvectors via asymptotic boundary conditions. In practice, it is used the Thomas-Fermi approximation for the kinetic term $\tau[n]=n^{2/3}/2$ while the exchange-correlation potential $v_{xc}[n]$ is parametrized as in \cite{PhysRevB.78.155412}.
	
\begin{figure}
		\centering
		\includegraphics[width=0.95\linewidth]{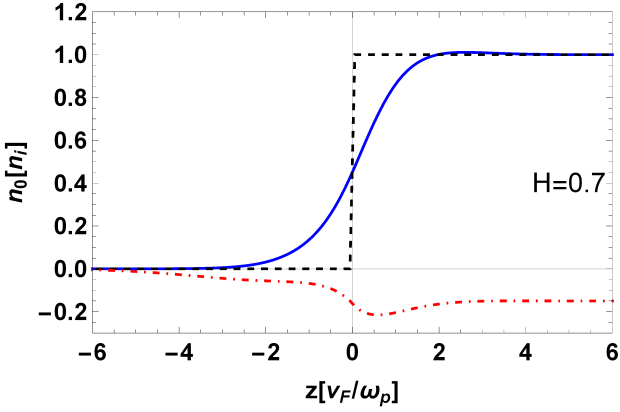}
		\caption{The numerical solution in the case $H=0.7$ for the ground state density $n_0(z)$ is represented in blue (solid line) while the step ionic background serves as reference being plotted with black (dashed line). Supplementary, the effective potential $g[n_0]+\varphi$ is shown in red (dot-dashed line).}
		\label{Fig_1}
	\end{figure}

	The QPHS is spatial invariant in the $Oxy$ plane, therefore, eq. \ref{eq_3} is solved only in the $Oz$ direction. An equidistant grid together with a finite difference scheme are used to represent the operator $\hat{h}$ on the $[-L,L]$ interval with $L\gg 1$ in order to reproduce the asymptotic conditions $\lim\limits_{z\to\infty}\psi(z)=1$ and $\lim\limits_{z\to-\infty}\psi(z)=0$. The ground state solution is obtained iteratively starting with a guess for $\psi$, computing at each step the effective potential $g[n]+\varphi$ from the eigenvector with the lowest energy and diagonalizing $\hat{h}$. A typical result for $H=0.7$ is shown in Fig. \ref{Fig_1} with an Wood-Saxon like profile for the equilibrium density $n_0(z)$, confined by the effective potential $g[n]+\varphi$ (while $\tau>0$, $v_{xc}<0$ and $\varphi<0$).
	
	In order to investigate the SP phenomena, we go back to the QHM and consider small time dependent fluctuations of the density and velocity field around their equilibrium values $n=n_0+n^1$ and $\mathbf{u}=0+\mathbf{u}^1$. The system of eqns. \ref{eq_1}-\ref{eq_2} is linearized and the velocity field is eliminated leading to a single equation for $n^1$:
	\begin{eqnarray}\label{eq_4}
		\left.\partial_{tt}n^1=\nabla\{n_0\nabla[\varphi^1+\frac{\delta g}{\delta n}\right|_0n^1-\frac{H^2}{2}Q^1]\}
	\end{eqnarray}
	where $Q^1$ is the linearization of the Bohm term $n^{-1/2}\nabla^2n^{1/2}$ and $\nabla^2\varphi^1=-n^1$. Using the the spatial invariance in the $Oxy$ plane ($\mathbf{r}_\parallel$), we move in the frequency-momentum domain applying \cite{Moradi} the time-space Fourier transform $\tilde{n}(z,q,\omega)=\int dtd\mathbf{r}_\parallel exp(-i\mathbf{q}\mathbf{r}_\parallel+i\omega t)n^1(\mathbf{r}_\parallel,z,t)$ on eq. \ref{eq_4}. After differentiation and algebraic manipulation we obtain the following :
	\begin{eqnarray}
		\label{eq_5}\mathcal{\hat{L}}\tilde{n} &=& \omega^2\tilde{n}\\
		\mathcal{\hat{L}}=\sum_{i=0}^4a_i(z,q)\partial_z^{(i)}&+&\frac{2\pi}{q}n_0'\partial_z\int dz'e^{-q|z-z'|}\label{eq_6}
	\end{eqnarray}
	An eigenvalue problem has been derived for the density modes $\tilde{n}$ with $\omega^2$ eigenvalues for the linear operator $\hat{\mathcal{L}}$. The latter is a fourth order differential operator with variable coefficients completed by the integral nature of the Coulomb term. The explicit expressions for $a_i(z,q)$ coefficients (presented in the Appendix\ref{appendix} \ref{eq_11}) are overwhelmingly complex making any analytical result almost unachievable.
	
	From a numerical perspective, $\hat{\mathcal{L}}$ is represented also within a finite difference scheme on the same $[-L,L]$ grid, supplemented by null Dirichlet boundary conditions at $z\to\pm L$. Diagonalizing the resulting matrix, the eigenmodes $\tilde{n}_j$ and associated eigenvalues $\omega_j$ are obtained. A first issue is that the dimension of the spectrum will be equal with the number of points on the grid, therefore, some supplementary criteria are needed to extract only the physical modes which are associated with the SP.  
	
\begin{figure}
		\centering
		\includegraphics[width=0.95\linewidth]{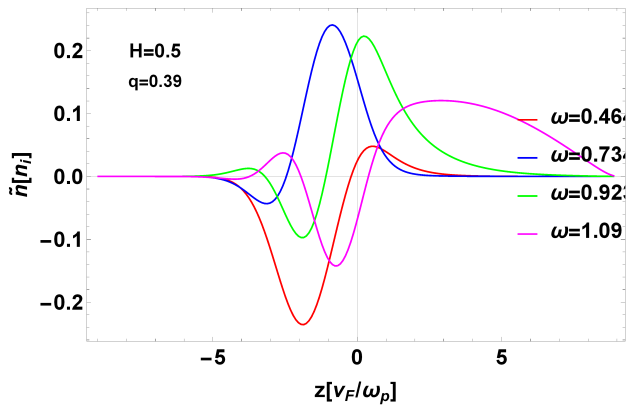}
		\caption{In the case $H=0.7$ and $q=0.39$ the first $5$ density modes $\tilde{n}(z;q\omega)$ are plotted. The fifth mode has an eigenvalue $\omega=1.12$ slightly larger than the critical $\omega_c=1.10$, thus becoming a bulk mode.}
		\label{Fig_2}
	\end{figure}

	Such a criteria is given by the asymptotic behaviour at $z\to\infty$, where $n_0(z)\to1$ and $n_0'(z)\to 0$. In this bulk region, $\mathcal{\hat{L}}$ simplifies to:   
	\begin{equation}\label{eq_7}
		\mathcal{\hat{L}}=\frac{H^2}{4}\partial_z^{(4)}-(\frac{H^2q^2}{2}+s_0)\partial_{zz}+1+\frac{H^2q^4}{4}+q^2s_0
	\end{equation}
	where $s_0$ is defined in the Appendix. The local solution for the associated differential equation is given as a superposition of $\exp(\pm \gamma_\pm z)$ terms where (similarly with \cite{Moradi}):
	\begin{equation}
		\label{eq_8}
		\gamma_\pm=k^2+\frac{2s_0}{H^2}(1\pm\sqrt{1+\frac{H^2}{s_0^2}(\omega^2-1)})\end{equation}
	Since by definition the SP must be localized at the surface of the quantum plasma with a decaying behaviour in the bulk region, then there is a critical value for the frequency $\omega<\omega_c$ (above this value, a surface mode turns into a volume plasmon) given by:
	\begin{equation}\label{eq_9}
		\omega_c^2=1+q^2(s_0+\frac{H^2}{4})
	\end{equation} 
	Performing extensive variation of the $(H,q)$ parameters it was found the existence of a variable number of modes that obey the critical condition. For small $H\approx 0.1$ we have $6$ modes while at the other end, $H=1$, there are only $4$ modes. In Fig. \ref{Fig_2} are plotted the spatial profiles of the first five $\tilde{n}_j(z;q,\omega_j)$. It can be observed that the fifth profile is associated with a frequency slightly larger than $\omega_c$ and for that, it does not show the searched decaying behaviour. Qualitatively similar results can be found for any $(H,q)$ pair, with surface plasmons of various localizations.
	
	Beside this cutting condition, it must be noted that $\mathcal{\hat{L}}$ is a non-hermitian operator which usually means complex spectrum and non-orthogonal eigenvectors. Despite this, all the surface modes have been found to be real by means of numerical simulation, property which is most likely connected with the fact that the finite difference matrix representation of $\hat{\mathcal{L}}$ is diagonally positive dominant. Further, due to non-orthogonality of the densities $\tilde{n}_j$, we face a spectrum of non-normal modes. The main physical consequence is that even exciting a single surface mode, all others will be automatically  excited due to their non-zero overlapping. For that, a supplementary criteria to evaluate the relative importance of each eigenmode during the realistic dynamics of a quantum plasma is needed.  
	
\begin{figure}
		\centering
		\includegraphics[width=1.\linewidth]{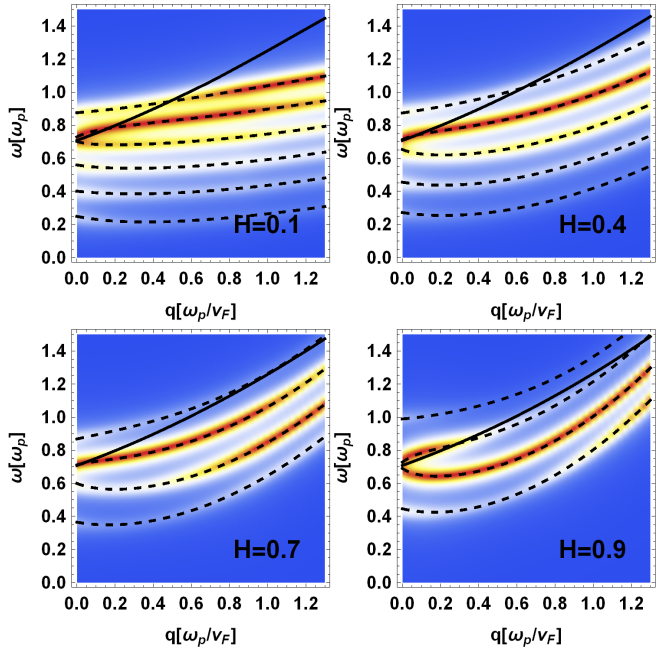}
		\caption{The spectrum function $\mathcal{S}(q,\omega)$ is plotted in coloured gradient. For comparison, the dispersion relation from \cite{Moradi} is shown in solid line, while the present eigenvalues are plotted with dashed line.}
		\label{Fig_3}
	\end{figure}

	Let us consider that the SP in a QPHS is excited by electrostatic interaction with a fast passing external point charge (other excitation mechanisms are possible but this was chosen due to its relevance for EEL experiments). In this case, the external field can be modelled as $\varphi_{ext}=\delta(t)e^{-qz}e^{i\mathbf{q}\mathbf{r}_\parallel}/q$. Including this potential in the momentum eq. \ref{eq_2}, an initial velocity field $\mathbf{u}(z,t=0)=e^{-qz}\mathbf{e}_z$ is obtained. Therefore, the relative importance of a certain eigenmode it is given by the overlap between the induced velocity field and the velocity field associated with $\tilde{n}_j$. The latter can be also obtained from the linearized momentum eq. as $i\omega_j^{-1}\hat{\mathcal{M}}^{-1}\tilde{n}_j$, where $\hat{\mathcal{M}}n=s_0n-Q^1+\varphi$. In this respect we define the spectrum function of surface plasmons as:
	\begin{eqnarray}\label{eq_10}
		\mathcal{S}(q,\omega)=\eta\sum_i\frac{|\mathcal{G}(q,\omega_i)|^2}{(\omega-\omega_i)^2+\eta^2}\\
		\mathcal{G}(q,\omega_i)=\frac{1}{\omega_i|\tilde{n}_i|} \int e^{-qz}\hat{\mathcal{M}}^{-1}\tilde{n}_i dz
	\end{eqnarray}  
	Where $|n|=\int n(z;q\omega)dz$ is a normalization factor while $\eta\ll 1$ is a phenomenological damping coefficient. The full results of this spectrum are shown in Fig. \ref{Fig_3} for some values of $H$. In solid line is plotted for comparison the dispersion relation for the SP in a \emph{bounded} QPHS obtained recently by \citet{Moradi} who used the correct set of boundary conditions\cite{BC}. The eigenvalue dependence $\omega_j(q)$ is specified with dashed lines while the spectrum is represented through the colour gradient. 
	
	Let us discuss two particular cases of the present model. First, in the classical limit $H=0$ it is well known that the SP becomes non-dispersive, i.e. $\omega=1/\sqrt{2}$. Unfortunately, eq. \ref{eq_3} does not allow for a numerical solution at $H=0$, therefore, such limit is not representable within the present framework. Nonetheless, we see in Fig. \ref{Fig_3} that when $H$ is decreased, the $\partial\omega/\partial q$ slope decreases for all eigenmodes, so we expect that in the classical limit, all frequencies will become flat, non-dispersive. Moreover, the amplitude in the spectrum will become localized around the $\omega\approx1/\sqrt{2}$ mode, consistent with the classical case while all other modes will be spurious and physically insignificant.
	
	The second particular case is the long-wavelength limit $q\to 0$ which has been predicted \cite{PhysRevB.3.220,Bennet} to be insensitive to the equilibrium profile $n_0(z)$. Due to the direct mapping between $n_0(z)$ and $H$, we evaluate the commutator $[\hat{\mathcal{L}}(H_1),\hat{\mathcal{L}}(H_2)]\neq\hat{\mathcal{O}}$ in the $q=0$ case. Since the operator $\hat{\mathcal{L}}$ does not commute for two random values of $H$ it means that its spectrum is sensitive to $H$ and so to $n_0(z)$ even for $q=0$. On the other hand, $n_0(z)$ is just weakly dependent on $H$ and the latter appears as $H^2$ in $\mathcal{L}$ which for usual values $H<1$ provides a slowly varying dependence of $\hat{\mathcal{L}}$ with $H$. !!This explains why in the numerical simulations it has been found a linear dependence between frequency and Planck constant, roughly  $\partial\omega/\partial H\approx 0.3$. 
	
	Regarding the $H$ dependence, from Fig. \ref{Fig_3} one can see how all frequencies are monotonically increasing functions of $H$, $\partial\omega/\partial H>0$. The first effect pushes upward all eigenvalues, up to the $\omega_c$ limit where they become bulk modes. In turn, the number of eigenmodes is a decreasing function of $H$ such that, while in the semi-classical limit $H=0.1$ we have $6$ surface plasmons, at $H=1$ only 4 have survived. The momentum dependence is qualitative quadratic at large $q$, consistent with the results of \citet{Moradi,LAzar} for the bounded QPHS. Quantitatively, the frequencies and their slopes are significantly smaller leading to a considerable red-shift. 
	
	Very important is that for any $H$, the most relevant eigenmodes in the spectrum are the ones which for $q\to 0$ are the closest to the classical value $\omega=1/\sqrt{2}$. This behaviour explains why in realistic experiments at metal surfaces it is not observed such a rich spectrum of modes. Moreover, for small $q$, some eigenvalues have a negative slope ($\partial\omega/\partial q<0$) consistent with experimental \cite{PhysRevLett.64.44} and  DFT calculations  \cite{PhysRevB.36.7378} for metal surfaces.
	
	Finally, it must be noted that while the qualitative behaviour is quite robust, some particular aspects as the specific values of the frequency or the number of SP modes are highly sensitive to the spill-out region of density. More precisely, small variations in the decaying exponent of the equilibrium density can change significantly results, in general shifting the frequencies and amplitude of the spectrum. This feature can be explained mathematically from the highly non-linear character of the coefficients, especially $a_0(z,q)$ 
	
	In the present work it has been investigated the phenomenon of surface plasmons in a quantum plasma half-space in the absence of any external confinement. This condition gives a continuous profile for the equilibrium density, which, in the frame of the Quantum Hydrodynamic Model, leads to a tangled differential eigenvalue problem with variable coefficients. By means of numerical simulation and defining a spectrum function, it has been proven that the unbounded QPHS supports multiple non-normal surface plasmon modes of various relevance for dynamics. The non-orthogonality implies that one cannot excite a single mode, but rather a superposition of modes. With respect to the bounded case, the dispersion relations show a significant red shift a flatter profile. This has an important impact in designing future quantum electronic devices where, for a desired larger frequency, a better external confinement would be needed. 
	
	This work was partially supported by the Romanian Ministry of National Education by the contract PN 16 47 01 01 with UEFISCDI. 
	
	\section*{Appendix}
	$s_0=\delta g/\delta n|_0=3n_0^{-1/3}/5+\delta v_{xc}/\delta n|_0$, $s_1=\ln(n_0)'$, $s_2=n_0''/n_0^2$, $a_4=H^2/4$, $a_3(z)=-H^2s_1/2$ and:
	
	\begin{widetext}
		\begin{eqnarray}\label{eq_11}
			a_2(z,q)=-n_0s_0+\frac{H^2}{2}(-q^2+s_1^2/2-2s_1')\\
			a_1(z,q)=H^2(\frac{q^2s_1^2}{2}-\frac{9s_1^3}{4}+3s_1s_2-\frac{3s_3}{4})-2(n_0s_0)'+n_0's_0\\
			a_0(z,q)=n_0+q^2n_0s_0-(n_0s_0')'+\frac{H^2}{4}(q^4+9s_1^4-2q^2s_1^2+2q^2s_2-17s_1^2s_2+4s_2^2+5s_1s_3-s_4)
		\end{eqnarray}
	\end{widetext}

	\bibliographystyle{apsrev4-1}
	\bibliography{multipleSP}
\end{document}